  \def   \ni {\noindent}

  \def   \bsk {\vskip 15truept}

  \def   \newline {\hfil\break}

  \input psfig.tex
  \magnification=1000
  \hsize 5truein
  \vsize 8truein
  \font\abstract=cmr8
  
  \font\text=cmr10     
  \font\affiliation=cmssi10
  \font\author=cmss10
  
  \font\title=cmssbx10 scaled\magstep2

  \def\ref{\par\noindent\hangindent 15pt}
  \nopagenumbers
  \null
  \vskip 3.0truecm
  \baselineskip = 12pt
  
  {\title                              
  \ni 
  Scanning strategies for the Planck mission
  }                                    
  \bsk \bsk
  {\author                             
  \ni  
  J. Delabrouille$^{1,2}$, J.-L. Puget$^1$, J.-M. Lamarre$^1$ 
  and R. Gispert$^1$
  }                                    
  \bsk
  {\affiliation                        
  \ni 1 Institut d'Astrophysique Spatiale,  
  Univ. Paris XI, 91405 Orsay Cedex, France

  \ni 2 Enrico Fermi Institute, Univ. of Chicago, 5460 S. Ellis Ave., 
  Chicago, IL 60637, USA
%
%
  }                                    
  \bsk
  \bsk
  \baselineskip = 9pt
  {\abstract                           
\ni 

}                                      
\bsk
\baselineskip = 12pt
{\text                                 
\ni 1. INTRODUCTION

The selection of a scanning strategy for the Planck mission is an important 
part of the optimisation of the mission. The sky coverage and 
more generally the distribution of the total mission integration time over 
the sky depend on it. In addition, the ability to subtract systematic effects 
from the data might depend drastically on the scanning strategy, as has been
illustrated on specific examples by Wright (1996) and Tegmark (1997).

At the time of the writing of this paper, no scanning strategy
for Planck is fully specified yet, but constraints have been put
which result from an iterative investigation of possible
orbit for the spacecraft, of general observing strategy, 
and of the optical, mechanical and thermal
design. This first order optimization, performed during 
the preparation phase of the project, has led to
constraints specified in the Announcement of Opportunity by ESA.
The satellite, on its halo orbit around the Sun-Earth L2
Lagrange point, 1.5 million kilometers away from the Earth, will be 
spun to 1 RPM. The optical axis is offset from the spinning axis by an 
``opening angle''
$\Theta_{\rm o}$ between 70 and 90 degrees, so that the beam of different
detectors will scan the sky on large 
circles which exact angular radius will be $\theta_{\rm scan} = \Theta_{\rm o}
+ \delta\theta$, where the detector-dependent displacement 
$\delta\theta$ is set by the position of the 
relevant light collector with respect to the center of 
the focal plane, and ranges between $-2.5$ and
$+2.5$ degrees. Every 60 scans or so, the spin axis position will
be offset by a few 
arcminutes, and the beam of each detector will scan
repeatedly a new circle on the sky.

The set of successive directions of the spin-axis can be
optimized, within the constraint that the solar aspect angle (angle between
spin axis and antisolar direction) cannot exceed about
10 degrees, and the earth
aspect angle at the times of data dumping cannot exceed a limit set by the
size of the main lobe of the telemetry antenna.

For our purpose, candidate scanning strategies are fully described by 
a small number of free parameters, which are the trajectory of the 
spin-axis on the sky, the set of spin-axis positions on this trajectory, the
time spent at each spinning position, and the opening angle $\Theta_{\rm o}$.

In this paper we discuss the problem of the optimisation of the 
trajectory of the spin axis on the sky and the choice of the opening angle
$\Theta_{\rm o}$. There are two main classes of scan strategies: those for
which the spin-axis is always anti-solar, and those for which it is allowed
to move away from the anti-solar position. For the first class thermal and
sidelobe effects due to the sun will be minimal. The drawbacks, as we shall 
see, are a reduced ability to estimate and remove systematic effects, and
possibly missing polar caps on the sky coverage.

\vskip 0.5truecm
}
 {\text 
\ni 2. REQUIREMENTS

A good scan strategy should meet, as much as possible, the following
requirements:
\item A Redundancy: There should be enough redundancy to 
test and correct for the presence of systematic effects in the data
such as sidelobe stray signals or thermal low frequency drifts, in order
to obtain after processing the cleanest maps possible and reliable error
estimation.
\item B Minimization of systematics: As much as possible the scan
strategy should minimize the level of signal contamination by such 
systematic effects in the data streams themselves.
\item C Sky coverage: As much as possible,  Planck should provide full sky 
maps. 
\item D Robustness: The scan strategy should be such that the inversion of
the data (i.e. obtaining full sky 
maps from the data streams and useful cosmological
and astrophysical information from the maps) is possible even if a few days
of data are lost or if one detector fails during the mission.
\item E Adaptability: One should be able if necessary to select the best 
scan strategy 
in the light of the information gathered during the verification phase
after injection at L2.
 
\vskip 0.5truecm
}
 
{\text 
\ni 3. SCANNING STRATEGY, DATA STREAMS AND SKY MAPS
\vskip 0.2truecm

\ni 3.1. Systematics and sky maps
\vskip 0.1truecm

We first investigate the reprojection of scan-synchronous
systematic effects on final maps of the sky. 
In the near-ideal situation where all instrumental
uncertainties come from scan-synchronous systematics and pure white noise
(very optimistic indeed!),
data reduction can be trivially simplified to the simple problem of 
optimal reconnection of rings (obtained from averaging 60 scans or so)
into a map of the sky. The data on each
ring $i$ can be modeled as:

$$d_i(\phi) = u_i(\phi) + v_i(\phi) + n_i(\phi)$$
where $d_i$ is the data, $u_i$ the useful astrophysical signal from main
beam, $v_i$ the systematics and $n_i$ noise on ring $i$, all functions of 
angle $\phi$ on that ring. In a discrete version, the angle $\phi$ along
ring $i$ is binned in ``pixels'' so that $\phi$ takes discrete values. 

The scan-synchronous systematics can be decomposed on a basis of orthogonal
functions, for instance Fourier modes:

$$v_i(\phi) = O_i + \sum_m C_{i,m} \cos(2\pi m \phi)
+ \sum_m S_{i,m} \sin(2\pi m \phi)$$
where $m$ ranges from 1 to some maximum value $m_{\rm max}$.

For map making with proper estimation and correction of systematic
effects it is necessary to invert the data set and solve for both the useful signal (temperature on the sky) and parameters $O_i$, $C_{i,m}$ and $S_{i,m}$
describing the systematic effects. This is possible only if the system is
non-degenerate.
For a scan strategy
with anti-solar spin axis and $\Theta_{\rm o}= 90^\circ$,
the data set for any detector located at the center of the focal plane
would consist of great circles crossing at the south (SEP)
and north (NEP) ecliptic poles. Taking the origin for angle $\phi$ on 
each ring at the NEP, it is clear that constraints, which come only from
ring intersections at the poles, cannot permit to
disentangle systematic effects as for each ring we only constrain the linear
combinations:

$$O_i + \sum_{m} C_{i,2m} \hskip 1.5truecm {\rm and}
\hskip 1.5truecm  \sum_{m} C_{i,2m+1}$$
with no constraint on the $S_{i,m}$. If there is any scan-synchronous
systematic effect, it can not be estimated from the signal, nor removed,
and would reproject on the sky as vertical stripes along great circles
joining NEP to SEP.

In principle, the degeneracy is broken by the physical extent of the 
focal plane, as not all detectors in a given channel scan the 
sky on the exact same rings. 
Even so, the intersections providing the redundancy
from which systematics can be estimated are very localised around the poles,
at angles $\phi \simeq 0$ and $\phi \simeq \pi$ along the rings,
where the leverage on sine modes is small and low $m$ cosine modes form
a nearly degenerate system. This near-degeneracy extends to Fourier modes
of the signal from the sky along rings.

It should be clear that this degeneracy is independent on the modeling of 
the problem. If an other basis of functions (instead of Fourier modes) had
been used, different combinations of parameters would have been
degenerate (they can be obtained from those on Fourier modes by expanding
each Fourier mode on the set of new functions and replacing in the above 
linear combinations).

The conclusion is that if there are scan-synchronous, ring-dependent
systematic effects in the data, an antisolar spin axis with large
opening angle ($\Theta_{\rm o}$ close to 90 degrees) is not acceptable for 
Planck. 

\vskip 0.2truecm

\ni 3.2 Low-frequency drifts
\vskip 0.1truecm

Even if there are no scan-synchronous systematics in the data, there
will be spurious low-frequency drifts due to instabilities in the 
electronics (yielding the so-called $1/f$ noise), or to random 
thermal drifts of
elements of the spacecraft radiatively coupled to the detectors, for instance.
Such drifts can typically be modeled as a Gaussian random process $n(t)$ to be 
added to the data, which spectrum $S_n(f)$ is known to reasonable accuracy.
We put into $n(t)$ only that part of these drifts that cannot be estimated
and removed from the data using additional information from thermometers,
sensors, ... 

The way this kind of noise reprojects on rings of data obtained from averaging
consecutive scans is discussed in Delabrouille, G{\'o}rski and Hivon (1998).
All the noise power in a band of width $~f_{\rm spin}/N$ centered on some
harmonic $f_m = m f_{\rm spin}$ of the spinning frequency contributes to the
variance ${\sigma_m}^2$ of the noise in mode $m$ (the width of the band is 
set by the 
resolution in frequency, which is the inverse of the total duration of the
useful signal, $N T_{\rm spin}$).

It is a good approximation, for most noise spectra and most methods to 
recombine $N \gg 1$ scans into one ring,
to relate the variance ${\sigma_m}^2$ in each Fourier mode
of the residual noise on the ring to the spectrum $S_n(f)$ of the noise
on the data stream by:

$$ {\sigma_m}^2 = {{S_n(m f_{\rm spin})}\over{NT_{\rm spin}}} $$

Low frequency noise thus reprojects as low-$m$ Fourier modes on Planck rings,
and can induce striping on the maps. These low-$m$ modes can be estimated and 
removed from the data as can be done for scan-synchronous systematics 
(see Delabrouille, 1998a) if the system that allows to estimate and remove them
is not degenerate. Therefore, if there is significant low frequency noise at 
frequencies larger than the spinning frequency of the satellite, 
an antisolar spin axis with large
opening angle $\Theta_{\rm o}$, again, is not acceptable for Planck.

  \midinsert
  \vskip 0.5truecm
  \par\noindent
  \centerline{\psfig{figure=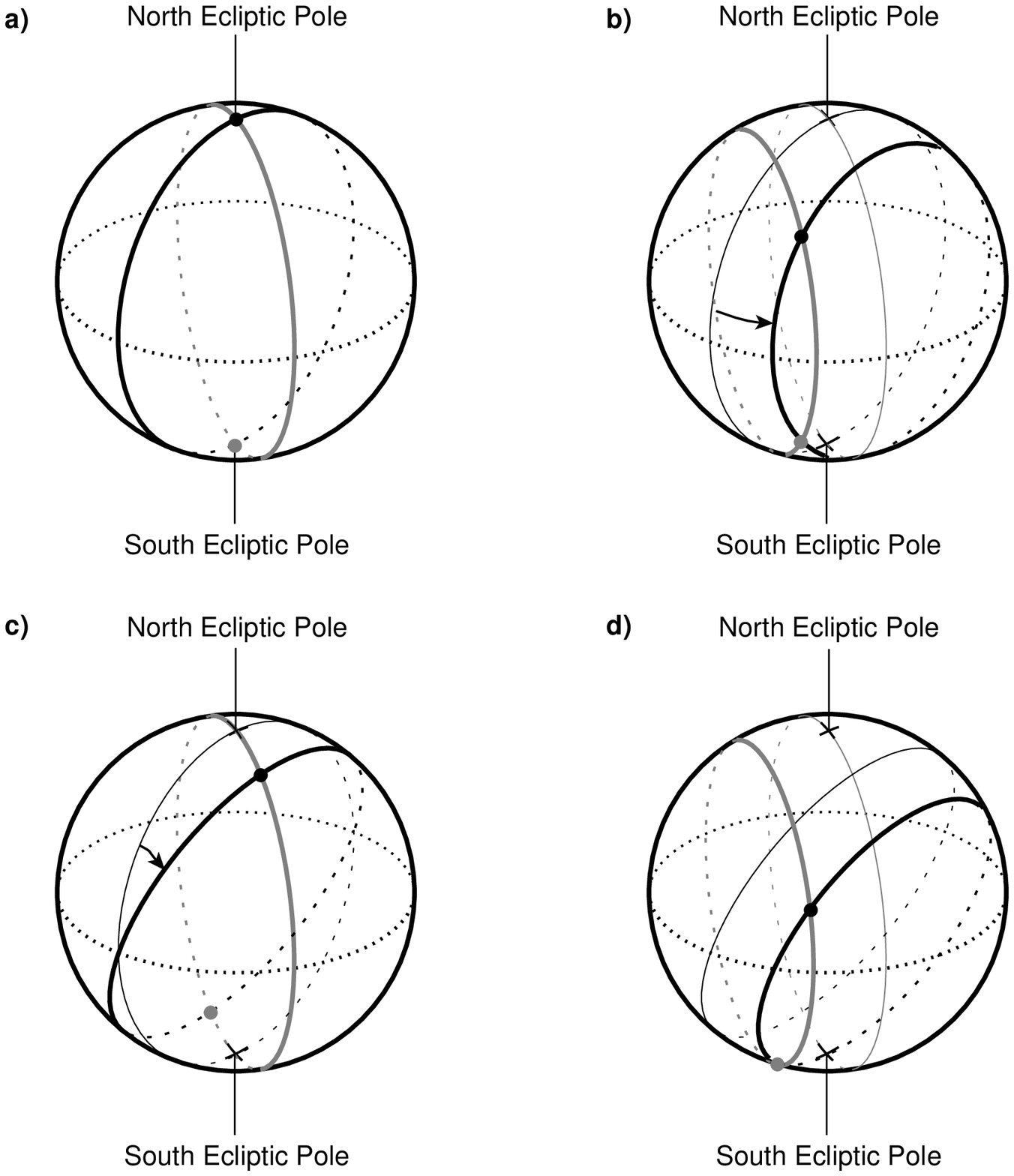,
                     width=10truecm}}
  
  \leftskip=1truecm
  \rightskip=1truecm
  \noindent
  {\abstract FIGURE 1.
  For $\Theta_{\rm o}=90^\circ$ and anti-solar spin axis at L2, any two Planck
  rings cross at the SEP and NEP (for a detector located on the optical axis),
  as illustrated on panel a. Panel b shows how reducing the opening angle
  changes the way Planck rings intersect on the sky for the same spin axes.
  Panel c shows the effect of allowing out of ecliptic motion of the spin
  axis (the spin axis for the thick black ring is now below the ecliptic 
  plane. Panel d, finally, illustrates how ring intersections depend 
  simultaneously on both spin-axis position and opening angle.    
   }
  \vskip 0.5truecm
  \leftskip=0truecm
  \rightskip=0truecm
  \endinsert

\vskip 0.2truecm

\ni 3.3. Minimizing systematics 
\vskip 0.1truecm

A solution to solve the degeneracy problem discussed above is to change the
solar aspect angle (angle between spin-axis and antisolar direction) during
the mission to guarantee better ring interconnections, as illustrated
in figure 1. It is clear,
however, that such an approach is likely to result in solar aspect 
angle dependent scan-synchronous effects during
the spinning of the spacecraft, induced by the pickup of 
scan-synchronous variations of the 
temperature of the payload, and by
modulated sidelobe signals from the Sun -- at the very least.
An anti-solar spin axis minimizes such effects. There is thus a trade-off
between redundancy for removing systematics by data processing, and 
minimization of scan-synchronous effects.

In order to make progress towards the optimal solution, we push the analysis
a bit further:
for a given solar aspect angle $\theta_{s,i}$ on ring $i$, the sun-induced 
scan-synchronous effect can be simply modeled as:

$$v_i(\phi) = O(\theta_{s,i}) + \sum_m C_m(\theta_{s,i}) 
\cos(2\pi m (\phi-\phi_i)) + \sum_m S_m(\theta_{s,i}) 
\sin(2\pi m (\phi-\phi_i))$$

This decomposition, which looks very similar to the Fourier expansion of 
paragraph 3.1, differs
by the fact that there is not a set of independent constants 
$(O, C_m, S_m)$ for each ring $i$, but one for each solar aspect angle.
It should be valid if there are no long transients in the temperature
evolution of the payload that depend on the spacecraft attitude history on
timescales of the order of one hour or larger (period between displacements
of the spin axis direction), or if these transients have only second-order
effects on the 1 minute period fluctuations and its harmonics. 
Angle $\phi_i$ (for each
ring $i$) is a phase, equal (for instance) to the angle $\phi$ for which
the main beam has the closest approach to the Sun (equal to 
$\pi - \theta_{\rm scan} - \theta_{s,i}$ for ring $i$).
If many circles are scanned with the same solar aspect angle, the number 
of parameters
needed to fit and remove the systematic effect is small, and if intersections
between rings provide a good connectedness, removing the systematic effect
should be easier.

It should be noted that thermal fluctuations of the payload will not be 
caused exclusively by the Sun: heat input due to the cycle of the sorption 
coolers, for instance, may generate significant temperature fluctuations 
of parts of the spacecraft, whose effect on the signal by pickup of
corresponding stray radiation cannot be 
modelled as a random noise, nor as scan-synchronous systematics.

The optimal scan strategy thus depends on the exact properties of all such
systematics and requires further studies, currently under way.

\vskip 0.2truecm
\ni 3.4. Sky coverage
\vskip 0.1truecm
 
A permanently anti-solar spin axis has the drawback that if the scan angle 
$\theta_{\rm scan}$ is different from 90 degrees for a given detector, 
polar caps around ecliptic poles are not covered with this detector. 
For Planck, even if the opening angle $\Theta_{\rm o}$ is
90 degrees, the scan angle will be smaller for some detectors because of
different locations in the focal plane. For some detectors and even some 
channels of Planck, small regions at the pole at least
a few degrees in diameter (depending on the value of $\Theta_{\rm o}$) 
would not be covered (for a given detector, 0.85 to 2.5 per cent of the 
sky is not covered for 
$\Theta_{\rm o} = 80^\circ$, depending on the location of the detector 
on the focal plane). Although this is not critical for 
characterizing the properties
of Cosmic Microwave Background anisotropies, it is not particularly desirable
either and should be kept in mind. 

Another aspect related to sky coverage is the distribution of integration
time over the sky. Maps of integration times for a selection of scan
strategies can be found in the COBRAS/SAMBA report on the phase A study
(Bersanelli et al., 1996) and in (Delabrouille, 1998b).

\vskip 0.2truecm

\ni 3.5. Robustness 
\vskip 0.1truecm

The full observing strategy is reasonably robust as there are several 
independent detectors in each channel and the sky is expected to be 
covered completely twice during the mission. It is highly recommended, 
however, that it be possible to obtain maps from single
detector data (at least for temperature measurements). If not,
the final sensitivity may be degraded if detector noise levels or 
contamination by systematic effects is very detector-dependent.
In addition, the ability to intercompare maps obtained with different
detectors will be a useful check for consistency in the data.

\vskip 0.2truecm

\ni 3.6. Adaptability
\vskip 0.1truecm

The optimal solution for a scanning strategy depends on the 
importance of scan-synchronous systematics for various strategies and
on the ability to process the data in order to clean the maps from such
spurious signals. An accurate evaluation of the level of residual
instrumental effects on maps after processing is difficult, as the 
estimation of systematics in various configurations depends 
sensitively on many instrument and mission parameters, which are not fully
known yet. For instance, thermal tests performed before launch and
mathematical models will be representative but probably not accurate enough to 
decide on the largest acceptable solar aspect angle. 
In addition, the development of optimal data processing 
algorithms is still in its infancy. It would not be wise, therefore, to
fully specify a scanning strategy at this stage.

Simulations can be helpful to isolate a small number of
acceptable scanning strategies, each of which is optimal or nearly so
for some models of dominant systematic effects.
But as it is quite possible that some aspects of the
simulations will prove inaccurate or incomplete, it 
is very important that tests be performed after
the injection of Planck on orbit, during a verification phase which will
last a few weeks. Such tests will permit to evaluate the
level and properties of systematics, check for consistency with models and
simulations and, if no nominal strategy can be specified relying on
simulations alone, decide at the very beginning of
the scientific mission which one of a few pre-selected scanning strategies
is optimal. Such a procedure could be repeated after one full sky 
coverage, and a different scanning strategy selected for the second
sky coverage. This adaptability of the scanning strategy
is important to the success of the mission.

\vskip 0.2truecm

{\text 
\ni 4. CONCLUSION
\vskip 0.1truecm

In this paper, we have identified a set of requirements for the Planck 
scan strategy, listed in section 2. In section 3, 
we have shown that a scanning strategy with anti-solar spin axis and an 
opening angle $\Theta_{\rm o} \simeq 90^\circ$ is extremely dangerous for
Planck, as it would preclude the possibility to check for scan-synchronous 
systematic effects or remove the effect of low-frequency drifts from the
maps. Moving the spin-axis away from the anti-solar direction helps removing
degeneracies, at the price of a likely increase of the
level of scan-synchronous, solar-aspect-angle-dependent systematics. This
must be quantified in order to decide which is the optimal scan strategy for
Planck.

If tests show that moving the spin-axis significantly away from the anti-solar
direction is impossible, then 
a reduced opening angle to $\Theta_{\rm 0} < 90^\circ$ breaks the
degeneracy, at the price of reduced sky coverage. Unfortunately, the 
opening angle cannot be readjusted in orbit, and should therefore be
conservatively chosen so that data reduction is possible even for very
pessimistic predictions of levels of possible systematic effects.

It is hard to evaluate at this stage which solution is the best without a
precise knowledge of the properties of systematics in each configuration
and of optimal ways of processing the data. An acceptable subset of optimal
solutions for various properties of systematics can be identified by
numerical simulations using as an input optimistic and pessimistic
thermal models of the spacecraft, simulated noise and evaluations of 
antenna patterns. If models are unsecure, the final decision between
possible solutions should be made in orbit after tests which will provide
a better understanding of the behaviour of the instrument in various 
configurations.
}
\vskip 0.2truecm

{\text 
\ni ACKNOWLEDGEMENTS
\vskip 0.1truecm

We thank the Local Organizing Comittee (LOC) for all the efforts made to
organize a very pleasant and enjoyable meeting, and
Jean Kaplan for useful comments and his careful reading of the manuscript.
}  
\vskip 0.truecm
 
{\text 
\ni REFERENCES
\vskip 0.1truecm

\ref Bersanelli, M., et al., {\sl COBRAS/SAMBA report on the phase A study},
ESA report D/SCI(96)3
\ref Delabrouille, J., 1998a, {\sl A\&ASS}, {\bf 127}, 555-567
\ref Delabrouille, J., 1998b, {\sl PhD thesis}, 
available from the author upon request
\ref Delabrouille, J., G{\'o}rski, K.M. \& Hivon, E., {\sl MNRAS},
{\bf 298}, 445-450
\ref Tegmark, M., 1997, {\sl Phys. Rev. D}, {\bf 56}, 4514-4529
\ref Wright, E.L., 1996, paper presented 22 november 1996 at the IAS
CMB data analysis workshop (astro-ph/9612006)
  }                     
  \end